\documentclass{aa}

\usepackage{multirow}
\usepackage{graphicx}
\usepackage{txfonts}
\usepackage{hyperref}
\usepackage[normalem]{ulem}
\usepackage[dvipsnames]{xcolor}
\hypersetup{
    pdftitle={Blazar PKS 0446+11},
    pdfauthor={Yuri Kovalev},     
    colorlinks=true,       
    linkcolor=MidnightBlue,
    filecolor=magenta,      
    urlcolor=violet,
    citecolor=teal,        
    pdfborder={0 0 0}
}
\usepackage{orcidlink}
\usepackage{xspace}

\graphicspath{{./}{figs/}}

\newcommand{\pks}{PKS\,0446+11\xspace}
\begin{document} 

\title{Multi-messenger flare in the quasar \pks}
\titlerunning{Multi-messenger flare in \pks}

\author{
Y.~Y.~Kovalev\inst{1}\orcidlink{0000-0001-9303-3263}
\and
M.~F.~Aller\inst{2}\orcidlink{0000-0003-2483-2103}
\and
A.~K.~Erkenov\inst{3}
\and
J.~L.~Gómez\inst{4}\orcidlink{0000-0003-4190-7613}
\and
D.~C.~Homan\inst{5}\orcidlink{0000-0002-4431-0890}
\and
P.~I.~Kivokurtseva\inst{6}\orcidlink{0000-0003-2955-3904}
\and
Yu.~A.~Kovalev\inst{7,6}\orcidlink{0000-0002-8017-5665}
\and
M.~L.~Lister\inst{8}\orcidlink{0000-0003-1315-3412}
\and
P.~V.~de~la~Parra\inst{9}\orcidlink{0000-0001-5957-1412}
\and
A.~V.~Plavin\inst{10}\orcidlink{0000-0003-2914-8554}
\and
A.~V.~Popkov\inst{11,7}\orcidlink{0000-0002-0739-700X}
\and
A.~B.~Pushkarev\inst{12,7}\orcidlink{0000-0002-9702-2307}
\and
A.~C.~S.~Readhead\inst{13}\orcidlink{0000-0001-9152-961X}
\and
E.~Shablovinskaia\inst{1}\orcidlink{0000-0003-2914-2507}
\and
Yu.~V.~Sotnikova\inst{3,6}\orcidlink{0000-0001-9172-7237}
\and
O.~I.~Spiridonova\inst{3}\orcidlink{0009-0007-7315-3090}
\and
S.~V.~Troitsky\inst{6,14}\orcidlink{0000-0001-6917-6600}
\and
V.~V.~Vlasyuk\inst{3}\orcidlink{0009-0002-6596-7274}
}

\institute{
Max Planck Institute for Radio Astronomy, Auf dem Hügel 69, D--53121 Bonn, Germany\\\email{yykovalev@gmail.com}
\and
Department of Astronomy, University of Michigan, 323 West Hall, 1085 S. University Avenue, Ann Arbor, MI 48109, USA
\and
Special Astrophysical Observatory of the Russian Academy of Sciences, Nizhny Arkhyz 369167, Russia
\and
Instituto de Astrofísica de Andalucía-CSIC, Glorieta de la Astronomía s/n, 18008 Granada, Spain
\and
Department of Physics and Astronomy, Denison University, Granville, OH 43023, USA
\and
Institute for Nuclear Research,
60th October Anniversary Prospect 7a, Moscow 117312, Russia
\and
Lebedev Physical Institute, 
Leninskiy prospekt 53, Moscow 119991, Russia
\and
Department of Physics and Astronomy, Purdue University, 525 Northwestern Avenue, West Lafayette, IN 47907, USA
\and
CePIA, Astronomy Department, Universidad de Concepci\'on,  Casilla 160-C, Concepci\'on, Chile
\and
Black Hole Initiative, Harvard University, 20 Garden St, Cambridge, MA 02138, USA
\and
Moscow Institute of Physics and Technology, Dolgoprudny, Institutsky per., 9, Moscow region, 141700, Russia
\and
Crimean Astrophysical Observatory, 298409 Nauchny, Crimea
\and
Owens Valley Radio Observatory, California Institute of Technology,  Pasadena, CA 91125, USA
\and
Physics Department, Lomonosov Moscow State University, 1-2 Leninskie Gory, Moscow 119991, Russia
}

\date{Received 6 November 2025; accepted 27 March 2026}

\abstract
{
The physical mechanisms driving neutrino and electromagnetic flares in blazars remain poorly understood.
}
{
We investigate a prominent multi-messenger flare in the quasar \pks to identify the processes responsible for its high-energy emission.
}
{
We analyze the IceCube-240105A high-energy neutrino event together with contemporaneous observations in the gamma-ray, X-ray, optical, and radio bands. The on- and off-flare spectral energy distributions (SEDs) are modeled within a single-zone leptohadronic framework. Multi-epoch VLBA observations from the MOJAVE program provide parsec-scale polarization data that complement the multi-wavelength light curves.
}
{
No significant time delay is detected between the neutrino arrival and the flares in different energy bands. This is consistent with an extremely small jet viewing angle below 1~deg, inferred from the parsec-scale polarization structure. The flare can be reproduced by the injection of a proton population and an increase of the Doppler factor from 18 to 24. We also detect an approximately 90~deg rotation of the EVPA in the parsec-scale core during the initial phase of the flare, indicating the emergence of a shock formed by the change in the bulk plasma speed.
}
{
Our comprehensive multi-messenger analysis demonstrates that the extreme beaming and sub-degree viewing angle of this distant blazar can account for the observed neutrino and electromagnetic activity. These findings strengthen the case for blazars as efficient accelerators of hadrons and as significant contributors to the observed high-energy neutrino flux.
}

\keywords{
Neutrinos --
Radio continuum: galaxies --
Galaxies: active -- 
Galaxies: jets -- 
Quasars: individual: \object{\pks}
}

\maketitle
\nolinenumbers

\section{Introduction}
\label{s:intro}

The growing dataset of high-energy astrophysical neutrinos, while still limited, provides unique opportunities to uncover connections with electromagnetic counterparts.
Various electromagnetic observational indicators have been employed to identify population-level associations between blazars and neutrinos, with bright compact radio emission proving most effective as an indicator of neutrino emission. This effect stems from relativistic beaming, where both radio emission and neutrinos are observed predominantly along the jet axis \citep{Plavin25}. Population studies have examined both time-averaged properties of compact radio emission and radio flares coincident with neutrino detections \citep[e.g.,][]{Plavin20,Hovatta21,Kouch24}.

Several individual blazars have exhibited clear electromagnetic flares coincident with high-energy neutrino events. TXS~0506+056 showed enhanced activity across the entire electromagnetic spectrum, from gamma-rays \citep{0506-scienceGamma} to radio wavelengths \citep{2019MNRAS.483L..42K}. The quasar PKS~1502+106 displayed a pronounced radio flare \citep{r:kielmann1502ATel} while remaining quiescent at other wavelengths, an association revealed through population analysis \citep{Plavin20}. The BL~Lac object PKS~0735+178 produced several neutrino events detected simultaneously by four observatories, accompanied by multiwavelength flaring \citep{2023ApJ...954...70A,2023MNRAS.519.1396S,2024MNRAS.527.8746P,2024MNRAS.529.3503B,2025ApJ...989..208P,2025A&A...699A.381K}. An untriggered flare search by ANTARES identified a coincident neutrino--gamma--radio flare in the quasar PKS~0239+108 \citep{2024ApJ...964....3A}. Additional blazars have also shown indications of enhanced electromagnetic emission temporally associated with neutrino detections \citep[e.g.,][]{2016NatPh..12..807K,giommi_3hsp_2020,liao_gb6_2022,2023IAUS..375...91E,2025ApJ...986..110J,2025A&A...698L...2P}.
Nevertheless, clear and statistically significant coincidences between high-energy neutrinos and broadband electromagnetic flares remain rare. This motivates our investigation of the recent multi-messenger event associated with the blazar \pks.

The IceCube alert IceCube-240105A was announced in GCN Circulars 35485 and 35498 \citep{2024GCN.35485....1I,2024GCN.35498....1I} as a track-like event of the BRONZE stream \citep[see a description of the alert system and types of the alerts in][]{IC_alerts,IceCat-1}.
This alert has an estimated false alarm rate of 2.64 events per year due to atmospheric backgrounds.
Epoch of arrival: 5~January~2024, 12:27:42.57~UT.
J2000 positions with uncertainties on the 90\,\% point spread function containment level, which include systematics: Right Ascension $72.69^{+1.92}_{-1.85}$~deg,
Declination $+11.42^{+0.50}_{-0.44}$~deg. 

These circulars have indicated two \textit{Fermi}-detected blazars in the error region. One of them was the low-synchrotron-peaked quasar \pks with a redshift 2.153 \citep{2012ApJ...748...49S}, and a bright parsec-scale jet.
Its J2000 VLBI sky coordinates are given by the Radio Fundamental Catalog \citep{2025ApJS..276...38P} as 
Right Ascension 04:49:07.671103, 
Declination $+$11:21:28.59628
with an accuracy of better than 0.2~mas,
catalog version rfc\_2025c.
Its angular distance to the event position is $0.40^\circ$.
The apparent VLBI speed of its relativistic jet was measured for two components in the units of the speed of light $c$ by \citet{2019ApJ...874...43L} to be $6.72\pm0.47$ and $6.48\pm0.76$.
Soon after the IceCube alert, many groups around the world announced the start of a major electromagnetic flare in \pks from gamma-rays to radio.
These include the following references:
\citet{2024ATel16398....1S,
Prince2024,
Woo2024,
Sharpe2024,
2024ATel16407....1G,
2024ATel16399....1E,
2024ATel16409....1K,
2025ATel17008....1V}.

\section{Observational data}
\label{s:data}

\subsection{The radio-to-gamma light curve and VLBA observations}

\begin{figure}
\centering
\includegraphics[width=\linewidth]{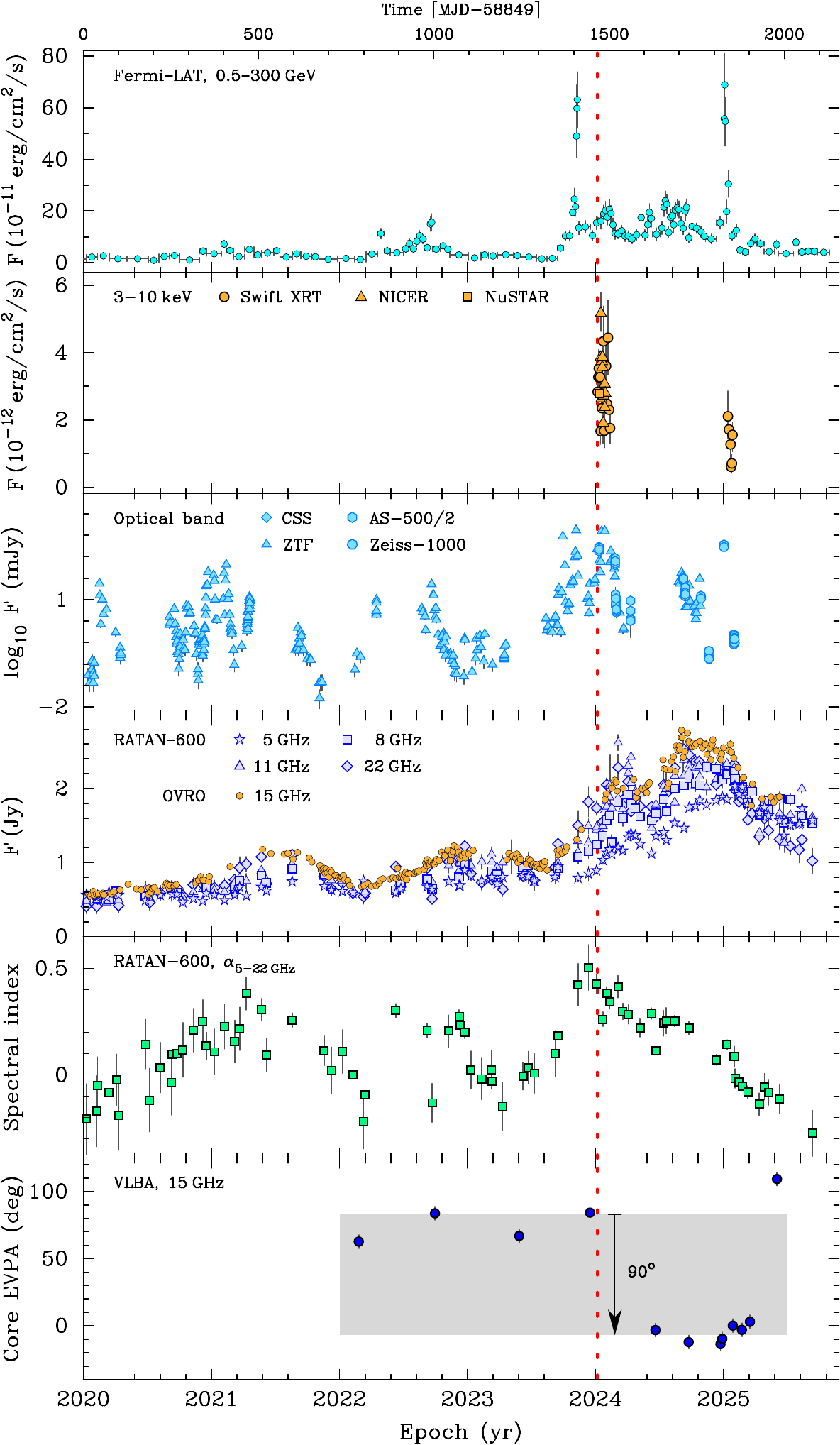}
\caption{
Multi-band light curve for \pks, shown for the time range around the high energy neutrino arrival time.
From top to bottom: 
gamma-ray, X-ray, optical, radio light curves.
They are supplemented by RATAN spectral index $\alpha$, calculated for $S\propto\nu^{+\alpha}$ between 5 and 22~GHz, and electric vector position angle (EVPA) of the parsec-scale core at 15~GHz, with a gray stripe of a $90^\circ$ width.
Uncertainties are shown as $1\sigma$ values.
The IceCube-240105A neutrino event epoch is shown by the vertical red dashed line.
See details in \autoref{s:data} and \autoref{sa:obs}.
\label{f:lc}
}
\end{figure}

\begin{figure*}
\centering
\includegraphics[width=\linewidth]{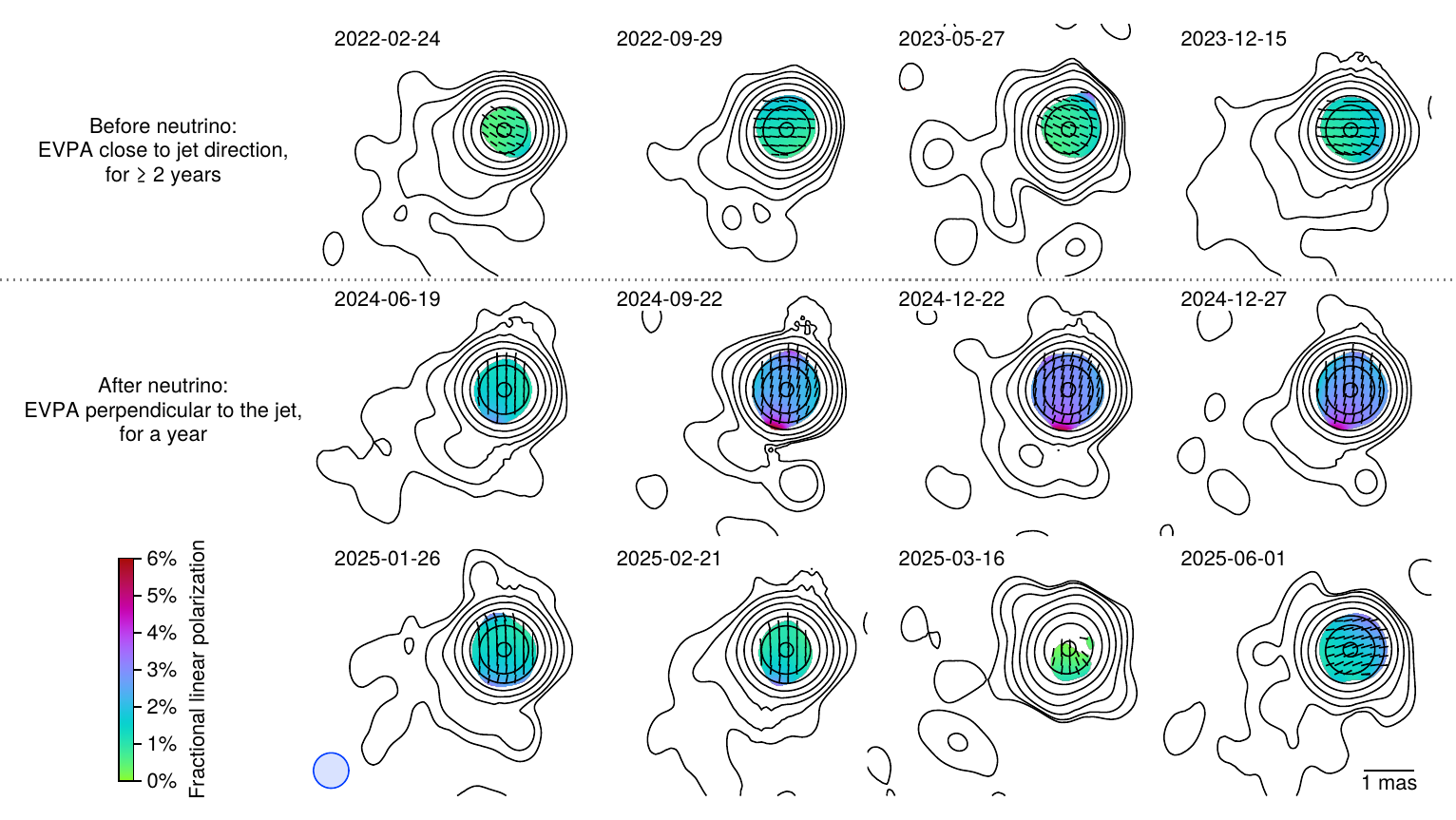}
\caption{
A set of polarization MOJAVE images, observed by the VLBA at 15~GHz. The images are reconstructed using the CLEAN method and the same circular restoring beam, its FWHM size of 0.7 mas is shown in the bottom left corner.
Stokes I is indicated by contours, with fractional polarization superimposed in false color. 
Stokes I contours start from a $4\sigma$ rms level.
The EVPA is indicated by sticks. Drastic changes in the polarization direction around the neutrino arrival (5~January 2024) are readily apparent, see also \autoref{f:lc}.
\label{f:MOJAVE_pol}
}
\end{figure*}

To analyze the variable electromagnetic emission before and during the flare associated with the high-energy neutrino event, we use both our own observations and reduced public observations of \pks across the electromagnetic spectrum. The combined light curve around the neutrino event appears in \autoref{f:lc}, while the full 20-year coverage is shown in \autoref{fa:lc}. A set of 15~GHz polarization VLBA images of this blazar from the MOJAVE program is presented in \autoref{f:MOJAVE_pol}.
We provide a detailed discussion of the observations, data processing, and calibration in \autoref{sa:obs}.

\subsection{A collection of calibrated data for SED}
\label{s:obs_public}

For the purpose of SED modeling (\autoref{s:SED}), we collect multiwavelength flux measurements making use of the Markarian Multiwavelength Data Center \citep[MMDC,][]{MMDC} data collection tool. These include measurements from VLA \citep{VLSSR,VLASS}, GLEAM \citep{GLEAM}, TGSS \citep{TGSS}, Green Bank Observatory \citep{NORTH20,GB87,GB6},  RACS \citep{RACS}, NVSS \citep{NVSS}, CRATES \citep{CRATES}, Planck \citep{PCNT}, ALMA \citep{ALMA}, WISE and NEOWISE \citep{WISE,NEOWISE,UnWISE}, 2MASS \citep{2MASS}, PanSTARRS \citep{Pan-STARRS,Pan-STARRS1}, SDSS \citep{SDSS}, Gaia \citep{Gaia}, ZTF \citep{ZTF}, UVOT \citep{UVOT}, ASAS-SN \citep{ASAS-SN}, Swift \citep{SWIFT-OUSXB,SWIFT-2SXPX,MMDC}, eROSITA \citep{eROSITA} and NuSTAR \citep{NuSTAR-NuBlazar}, together with data from \citet{Kuehr1981,SPECFIND,HSTGSC,UVOT}. 
For explanation of acronyms, see the cited papers.

We supplement the data provided by MMDC with more data from ALMA \citep{ALMA} and the data described in \autoref{sa:obs}. The data are split in two parts, ``flare'' (observation date 60200$\le$MJD$\le$60700) and ``off-flare'' (54684$\le$MJD$\le$60200 and historical catalog values). Individual observations at the same frequencies are averaged within these two time intervals, even if performed by different instruments. Fluxes are corrected for the Galactic extinction using the HI column density \citep{HI-column}, its relation to the V-band extinction from \citet{HI-to-extinction} and the color correction of \citet{Schlafly2011}. 

\subsection{Neutrino}
\label{s:neutrino}

The IceCube neutrino event associated with \pks was reported as an alert 240105A in the BRONZE stream \citep{2024GCN.35485....1I,2024GCN.35498....1I} with the estimated best-fit energy\footnote{\url{https://gcn.gsfc.nasa.gov/notices_amon_g_b/138821_46175426.amon}} of 109~TeV, while none were detected in the GOLD stream. The energy uncertainties may be estimated by scaling typical track alert energy errors \citep[see, e.g.,][]{0506-scienceGamma} to the best-fit energy of the present event; we obtain $109^{+755}_{-28}$~TeV. Note that for muon tracks which cross the neutrino detector volume, strongly asymmetric errors in the energy estimate are typical. For the flux estimate, we use the effective area of $\approx 20$~m$^2$, as reported for the combined GOLD and BRONZE sample in the online tables\footnote{\url{https://doi.org/10.7910/DVN/SCRUCD}} accompanying the IceCat-1 catalog \cite{IceCat-1} and interpolated to the best-fit neutrino energy of 109~TeV. Unfortunately the effective area is rather coarsely binned in declination (the bin we use is from 0$^\circ$ to $+30^\circ$). 

For the flare analysis, one event was used, and the Poisson confidence interval was adopted as $1_{-0.62}^{+1.35}$ at 68\,\% confidence level (CL). The neutrino flux is estimated as $(2.1^{+2.8}_{-1.3})\times10^{-11}$~erg\,cm$^{-2}$\,s$^{-1}$. 
This assumes the flare duration of 500~d, see the discussion in \autoref{s:SED} and \textit{Fermi} light curve analysis in \autoref{sa:flare}.
For the off-flare period, no neutrino alerts associated with \pks were present, and the Poisson upper limit of $<2.44$ events at 90\,\%~CL was used. This resulted in a neutrino flux limit of $3.0\times10^{-12}$~erg\,cm$^{-2}$\,s$^{-1}$. 

\section{A single-zone SED model}
\label{s:SED}

\begin{figure*}[t!]
\centering
\includegraphics[width=1.0\linewidth,trim=5cm 0cm 5cm 0cm,clip]{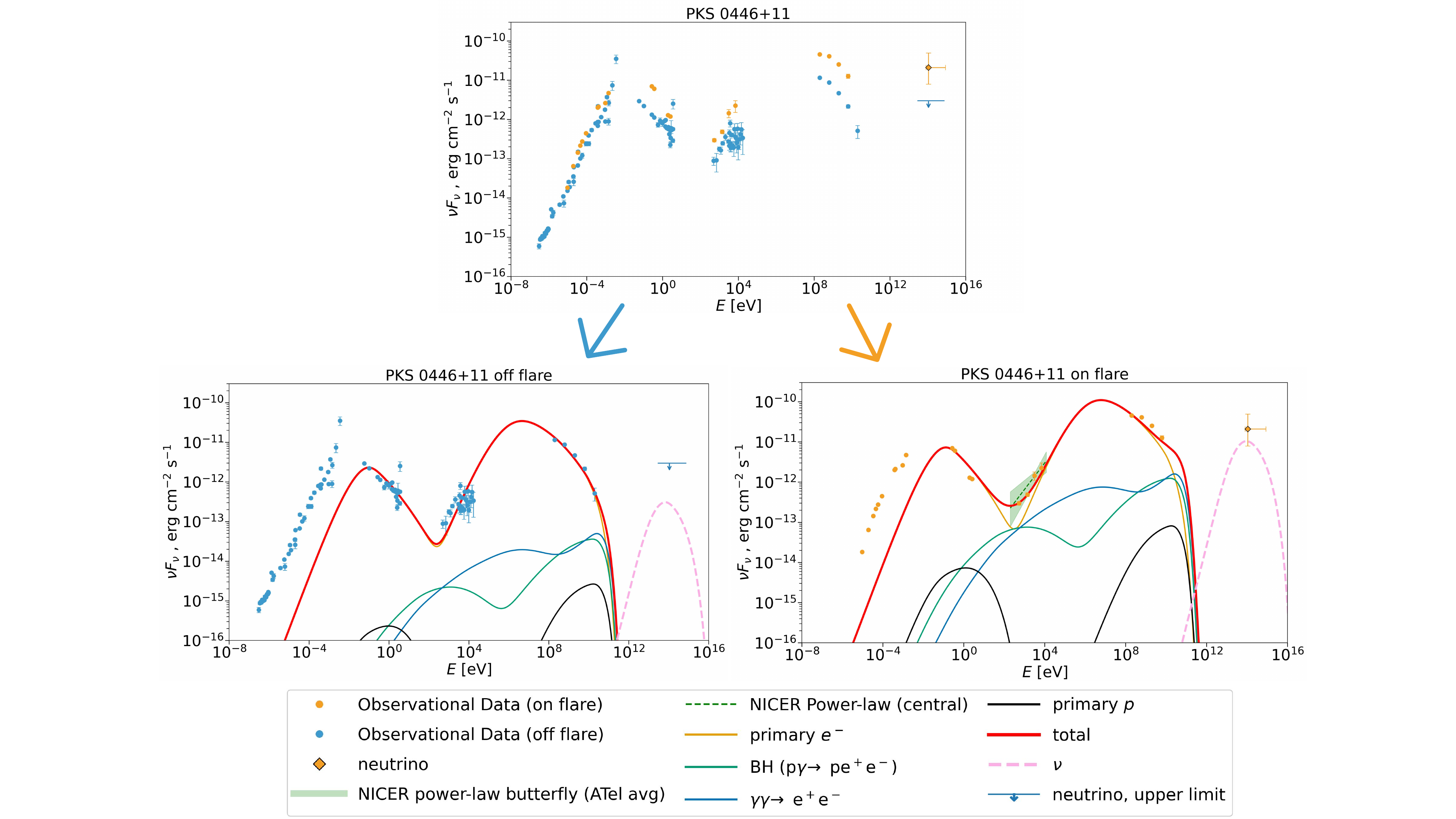}
\caption{
\textit{Top:} Combined Spectral Energy Distribution (SED) for the quasar \pks, \textit{bottom left:} the off flare period, \textit{bottom right:} the on flare period.  The total photon spectrum is shown as a thick red curve. The pink dashed line represents the neutrino spectrum (all flavors). The orange curve gives the sum of synchrotron and Compton emissions from injected electrons, while the black one is the synchrotron emission from injected protons. The green and blue curves present the secondary emission from $e^+e^-$ pairs generated via the Bethe–Heitler process and $\gamma \gamma$ annihilation, respectively. See the text for discussion of the observational data points.
\label{f:SED}
}
\end{figure*}

We use the data described above to obtain broadband SEDs of \pks for the off-flare and on-flare periods. These SEDs are presented in \autoref{f:SED}
together with predictions of a single-zone leptohadronic model, obtained with the help of the publicly available simulation code AM$^3$ \citep{AM3}. The parameters of the model are listed in \autoref{t:SEDparam}.  In our simulation, we adopt a baseline scenario and assume the same ballistic escape time for all particles, $t_\mathrm{esc}=R/c$.
Usually, SED modeling implies parameter degeneracy, so the solution presented here is a representative one, rather than unique.
We note that \citet{2026ApJ...999..245K} have independently modeled the SED for \pks, focusing on the period in early 2024.

The AM$^3$ code consistently takes into account standard electromagnetic processes: synchrotron emission, inverse Compton scattering and, where relevant, pair production, $\gamma\gamma \to e^+e^-$. Hadronic interactions with photons are modeled, including both those which lead to production of $\pi$ mesons, whose subsequent decays generate high-energy neutrinos and photons, and Bethe-Heitler pair production, $p\gamma \to p e^+ e^-$. For the purpose of the present work we neglect $pp$ interactions, whose probability is suppressed in blazar jets with respect to $p\gamma$ interactions. 

All secondary particles are tracked, so electromagnetic cascades initiated by the $e^+ e^-$ pair production and taking place in the source are followed down to photon escape energies. For the calculation of gamma-ray pair production on the extragalactic background light during their propagation from the source to the observer, we adopt the \citet{Gilmore_2012} fiducial model of the extragalactic background light (EBL). For the redshift of \pks, $z = 2.153$, the optical depth for gamma-ray energies $>100$~GeV is $\tau \gtrsim 2.6$ and rapidly grows with energy. This explains the sharp cutoff in the photon SED at these energies and makes it hardly possible to detect such energetic gamma rays from this source at the Earth, unless new physics affects the $e^+e^-$ pair production.

All processes relevant to both electromagnetic and neutrino emissions are assumed to take place in a spherical region, hereafter a blob, through which the relativistic plasma flow is passing, and where injected non-thermal electrons and protons interact. 
One may see from \autoref{t:SEDparam} that both flare and off-flare SEDs are described by very similar parameters, with the two key differences -- during the flare, both proton luminosity and Doppler factor increase. 

We verified that radiation fields external to the emission region have only a subdominant impact on the modeled SED. This includes the cosmic microwave background (CMB), accretion disk (AD), and broad-line region (BLR) emissions. 
The CMB contribution is negligible close to the jet base, where the photon field is dominated by the AGN emission. The influence of other external components depends on 
the location of the emitting blob. Assuming its preferred radius of $5\times10^{16}$~cm ($\approx 0.012$~pc), the blob may still be located farther from the black hole. The key question is whether it resides within or beyond the BLR. Including a noticeable BLR contribution worsened the fit compared to the baseline model. We therefore conclude that the neutrino-emitting region lies beyond the BLR, at distances $\gtrsim 0.1$~pc from the black hole, where BLR photons are deboosted and contribute insignificantly. 
This is supported by an independent beaming analysis of \citet{Plavin25}, who deduced that neutrino production happens at sub-parsec distances, where the jet is already partly accelerated. 
The same reasoning applies to AD photons. This interpretation is supported by the absence of a strong X-ray cascade component, which would otherwise arise from interactions with BLR and AD photons. The inferred larger distance is also consistent with the relatively low magnetic field suggested by our model.

The small size of the region, however, suggests that it is located close to the jet base and not far from the central black hole, $\lesssim 0.5$~pc, because it would expand adiabatically when moving to farther parts of the jet. This is supported by the fact that the SED does not describe radio measurements, for which the inner part of the jet is opaque. Data points in the radio band were used as the upper limit in our SED modeling, and the corresponding emission is assumed to come from different parts of the jet.

The estimate of the neutrino flux, which determines the hadronic part of the SED model, depends crucially on the assumed duration of the flare. Note that the inspection of the gamma-ray light curve, cf.~\autoref{f:lc}, reveals two timescales of the flare: short peaks of a few-day duration and the main flare of 500~d. 
For detailed analysis, see \autoref{sa:flare}.
If the neutrino emission were associated with the short timescale, the flux estimate would increase by two orders of magnitude, making the observed SED impossible to explain because of the required unrealistic proton power. We consider this as an additional argument that the neutrino emission is associated with longer processes, of the order of the duration of the main gamma-ray and radio flare.

\begin{table}
\caption{Parameters of the SED model}
\label{t:SEDparam}      
\centering                                     
\begin{tabular}{c c c}          
\hline\hline
Parameter  & On flare & Off flare \\
\hline
Doppler factor & 24 & 18 \\ 
Magnetic field, G& \multicolumn{2}{c}{0.06} \\ 
Blob radius, cm& \multicolumn{2}{c}{$5\times10^{16}$} \\ 
Min electron energy, eV&\multicolumn{2}{c}{$1.5\times10^{9}$} \\ 
Max electron energy, eV&\multicolumn{2}{c}{$9\times10^{10}$}\\ 
Min proton energy, eV&\multicolumn{2}{c}{$8\times10^{13}$}\\
Max proton energy, eV&\multicolumn{2}{c}{$6\times10^{14}$}\\ 
Electron power, erg/s&\multicolumn{2}{c}{$8\times10^{43}$}\\ 
Proton power, erg/s&$4\times10^{48}$& $4\times10^{47}$ \\ 
Electron spectral index&\multicolumn{2}{c}{3.4} \\
Proton spectral index&\multicolumn{2}{c}{1.9} \\
\hline                                            
\end{tabular}
\end{table}

The inferred parameters of the injected proton spectrum are quite unusual. The spectrum is parametrized with a segment of a power law, with the minimal and maximal energies being free parameters, together with the spectral index. We find that a hard proton spectrum is needed to describe all the data, and the difference between minimal and maximal energies is less than an order of magnitude. Attempts to describe the SED with a wider spectrum for protons fail because of an excessive contribution of secondary photons at the energies between the synchrotron and Compton bumps of the SED~--- this problem is also known for TXS~0506$+$056 \citep{0506-SED}. The most problematic contribution comes from electromagnetic cascades initiated by both Bethe-Heitler electrons and from energetic photons born in $\pi^0$ decays. 
Variants of proton acceleration mechanisms, possibly operating in the jet and providing such hard and narrow spectra, include magnetic reconnection \citep[e.g.,][]{GaoReconnection,NalewajkoReconnection}, the converter mechanism \citep{Converter}, transrelativistic shocks \citep{BykovJETP}, etc.
The observation of the high energy neutrino from \pks is a rare case where hints about the origin of accelerated protons may be obtained. This might be related to the extremely small viewing angle (\autoref{s:eye}), which opens access to a different combination of observables.

The on-flare value of the proton power injected in the neutrino emission zone, $L_p$, is an order of magnitude higher than the Eddington luminosity, $L_{\rm Edd}$, for a $10^9~M_\odot$ black hole. For a steady relativistic flow, the host-galaxy-frame jet power scales as $\Gamma^2$ times the comoving energy density. Thus, compared to the comoving proton-injection power, the corresponding host-galaxy-frame power budget is $\sim \Gamma^2 L_p$, up to a factor of order unity. Importantly, $L_{\rm Edd}$ is a limit on the radiative luminosity of quasi-spherical accretion, not a strict upper bound on the jet power which may be extracted from the black-hole rotation \citep{BZ}. Simulations of jet launching in this regime confirm that the jet power can greatly exceed $L_{\rm Edd}$ \citep[e.g.][]{Tchekhovskoy2011,Sadowski-Narayan}.
Observationally, kinetic jet proton power exceeding the Eddington luminosity by a factor of ten may even be common \citep{Ghisellini-power}. Upper limits on the proton power come instead from the secondary radiation from the associated electromagnetic cascades \citep[e.g.,][]{Murase-external,Reimer-external}; this cascade emission is accounted for in our SED model and does not overshoot the data. Therefore, for \pks, the required $L_p > L_{\rm Edd}$ does not present a problem.
The location of both the proton acceleration region and the neutrino production zone in the jet is consistent with observational results \citep[e.g.,][]{Plavin25,2025arXiv251016585K}, which indicate that neutrino blazars are preferentially selected through Doppler boosting.

\section{The 90-deg EVPA flip in the parsec-scale core}
\label{s:evpa}

The epochs before and after the neutrino event demonstrate a clear $90^\circ$ flip of the EVPA in the core region, as seen in the time series in \autoref{f:lc} and the polarization VLBI images in \autoref{f:MOJAVE_pol}. The polarization direction changes by about $90^\circ$ within six months or less after the neutrino arrival and flips back approximately one year later. We note that the uncertainties in the measured core EVPA are about $5^\circ$ or smaller.

The $90^\circ$ EVPA flip following the neutrino event may reflect (i) a substantial change in core opacity, (ii) the emergence of a shock, or (iii) an increase in its emission. 
\citet{2018Galax...6....5W} noted that a very high synchrotron opacity of $\tau > 6$ is required for the first scenario, which is observationally rare. Nevertheless, opacity-driven EVPA flips have been convincingly observed by \citet{2017Galax...5...81A,2022MNRAS.510.1480K,2024MNRAS.528.1697K}.

In the case of \pks, we concur with \citet{2018Galax...6....5W} and do not associate the EVPA flip with opacity effects, for the following reason. As seen in \autoref{f:lc} and \autoref{f:MOJAVE_pol}, the core EVPA remained around $+80^\circ$ for at least two years prior to the neutrino arrival, despite significant variations in the spectral index, which became inverted at the end of 2023 during the start of the radio flare.
After the neutrino arrival, the core EVPA shifted to about $-10^\circ$ and stayed there for nearly 1.5 years, while the radio flare exhibited two peaks and the spectral index varied between $+0.5$ and $-0.3$.
Since the spectral index variations are the most direct indicator of changing opacity, we conclude that the flip is not related to opacity variations. 
Moreover, the core fractional linear polarization increased after the flip (\autoref{f:MOJAVE_pol}), whereas a strong increase in opacity is expected to reduce the fractional polarization \citep[e.g.,][]{2017Galax...5...81A}.

We find that the emergence of a shock, or the enhancement of polarized emission associated with it, provides a more plausible explanation. 
\citet{1989ApJ...341...68H} showed that a shock can produce both a $90^\circ$ EVPA rotation and an increase in the fractional polarization, exactly as observed by us (\autoref{f:MOJAVE_pol}).
The polarized flux increases strongly after the neutrino arrival, likely dominating the original quiescent emission that may still persist in the unresolved core region.

\citet{2022ApJS..260....4P} estimated the jet direction to be at the position angle of $121^\circ\pm4^\circ$ at 15~GHz in the plane of the sky.
We note that the core EVPA values before and after the flip do not agree with parallel and perpendicular jet directions, as one could expect for a shock enhancement.
This can be explained by either the unknown Faraday Rotation Measure (RM of an order of one and a half thousand $\mathrm{rad\,m^{-2}}$ is sufficient) or a change in the inner jet direction. The latter is realistic to expect in projection, due to the low assumed jet viewing angle (see discussion in \autoref{s:eye}). 

\section{The jet viewing angle and Lorentz factor}
\label{s:eye}

\begin{figure}
\centering
\includegraphics[width=1.00\linewidth,trim=1.5cm 0.5cm 1.8cm 0.5cm,clip]{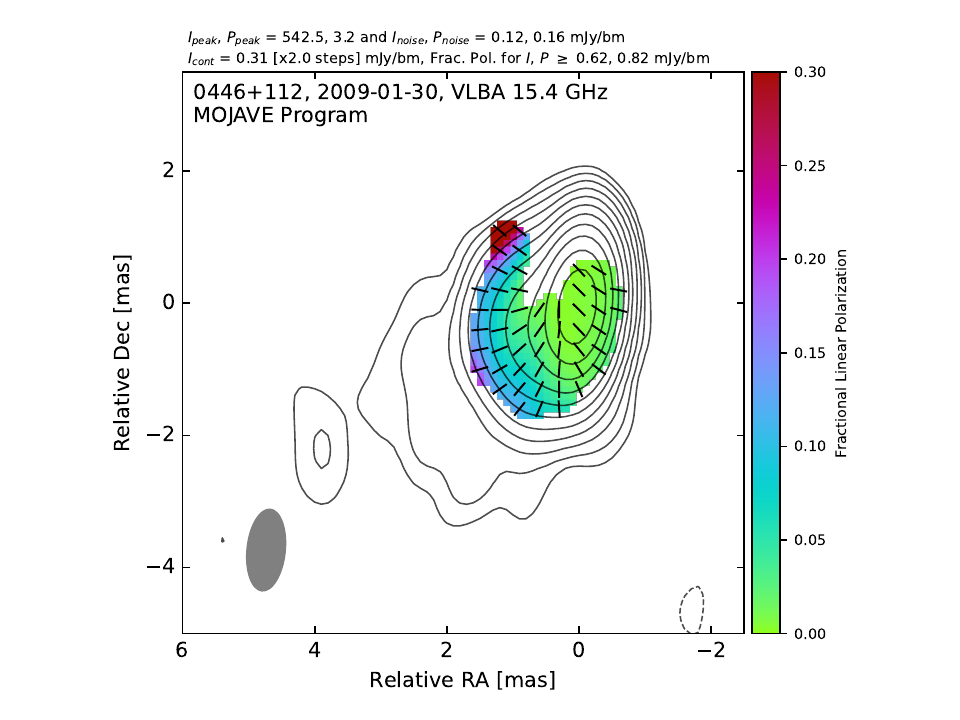}
\caption{
MOJAVE 15~GHz VLBA epoch with the weakest core emission.
Stokes I is shown by contours, while fractional linear polarization by false color and EVPA direction by sticks. Stokes I contours start from a $2.5\sigma$ level. The beam is presented at the Full Width at Half Maximum (FWHM) level as an elliptical Gaussian in the bottom left corner. 
\\
A radial pattern of EVPA directions is readily apparent, indicating close to a face-on viewing angle, see \autoref{s:eye}. This effect clearly seen in epochs with low core dominance because of dynamic range limitations: compare to \autoref{f:MOJAVE_pol}. 
Note that \autoref{f:MOJAVE_pol} color bar has a different scale, reflecting changes in the core polarization fraction over time.
\label{f:eye}
}
\end{figure}

For off-flare and on-flare blob Doppler factors $\delta=18~\&~24$ (\autoref{t:SEDparam}) and the apparent jet speed $\beta_\mathrm{app}=6.7\,c$ measured by MOJAVE \citep{2019ApJ...874...43L}, we obtain viewing angle estimates $\theta\approx2^\circ~\&~1.2^\circ$, respectively.
At the same time, $\beta_\mathrm{app}$ is measured at a projected distance of the order of $5$~pc from the apparent jet base. The single-zone emitting blob is expected to be located significantly closer to the central engine (see discussion above), where the plasma may still undergo substantial acceleration \citep{2015ApJ...798..134H,2020MNRAS.495.3576K} and the apparent speed is likely lower. 
Therefore, the values of $\beta_\mathrm{app}$ and $\theta$ quoted above should be regarded as upper limits.

Let us now consider the following.
The very high energy photon and neutrino emitter PKS~1424+240 was found to have a strong toroidal component in a face-on jet with a viewing angle about or less than $0.6^\circ$ \citep{2025A&A...700L..12K}. A radial EVPA pattern was also found by \citet{2002ApJ...580..742H} in a specific jet region of PKS~1510$-$089, observed face-on.
The EVPA directions in the jet of \pks show a similar radial pattern at one epoch, where the core emission doesn't dominate the jet, see \autoref{f:eye}. Together with its wide apparent opening angle of $42^\circ$ \citep{MOJAVE_XIV}, this strongly supports a very small jet viewing angle~--- smaller than its opening angle.

On this basis, we adopt $\theta = 0.6^\circ$, consistent with \citet{2025A&A...700L..12K}.
The Lorentz factor depends only weakly on the viewing angle in this regime and becomes $\Gamma\approx9~\&~12$ for the off-flare and on-flare periods, respectively.
We note that an increase in the plasma Lorentz factor may lead to the formation of an associated shock \citep[e.g.,][]{2001MNRAS.325.1559S}, discussed in \autoref{s:evpa}.

%

\section{Multi-messenger flare coincidence and the gamma-ray-to-radio delay}

Visual inspection of the light curve in \autoref{f:lc} reveals that the electromagnetic flare began a couple of months before the neutrino detection in all well-sampled bands (gamma-ray, optical, and radio). 
The parsec-scale core brightness also increases significantly already by the onset of the flare in late 2023.
This does not necessarily imply that the neutrino flare was delayed relative to the electromagnetic one. Given the low statistics of neutrino detections, the observations are compatible with a range of scenarios, from simultaneous flares across all bands and messengers to constant neutrino emission over time. Nevertheless, we interpret this as a simultaneous multimessenger flare in \pks.

Such a close temporal coincidence between the neutrino arrival and the broadband electromagnetic flare strongly suggests the neutrino emission is also flaring, making the situation qualitatively similar to that of TXS~0506+056.
A simple a-posteriori analysis yields a 2\,\% chance coincidence probability based on temporal information alone.
Among 314 VLBI-bright blazars (flux density $\geq150$~mJy in \citealt{2025ApJS..276...38P}) with radio and gamma-ray light curves available from OVRO and \textit{Fermi} LAT monitoring, only six (including \pks) exhibit gamma-ray and radio peaks within one year of the 5~January 2024 neutrino arrival.
See also \citet{2025RAA....25k5018C} for independent estimates.

The emission enhancement appears to begin simultaneously across all bands, from radio to gamma-rays (\autoref{f:lc}), with comparable timescales. Consequently, in \autoref{s:SED} we adopt the flaring scenario and assume that the neutrino flare timescale is similar to that of gamma-rays.
This behavior is unusual compared with major flares in other neutrino-candidate blazars, including TXS~0506+056 (see references in \autoref{s:intro}). The gamma-ray-to-radio lag $\Delta t$ \citep[e.g.,][]{2022MNRAS.510..469K} is typically related to  the synchrotron opacity \citep[e.g.,][]{2012A&A...545A.113P}: regions close to the jet origin are opaque to radio waves, but not to gamma rays. If the apparent base of a radio jet (radio ``core'') is located at a de-projected distance $L$ from the high-energy emitting blob (\autoref{s:SED}), the radio-to-gamma-ray delay is expected to be $\Delta t=L\sin\theta / \beta_\mathrm{app}$ from pure geometry. 
\citet{2012A&A...545A.113P} estimates the typical de-projected distance  between the true jet base and apparent 15 GHz parsec-scale core to be $L\sim10$~pc or $\sim30$~ly for the MOJAVE sample.

For approximate calculations, we adopt the jet viewing angle of $\theta = 0.6^\circ$ (\autoref{s:eye}). For the region of the high-energy blob close to the nucleus, we infer an apparent speed of $\beta_\mathrm{app} = 3c$, which corresponds to $\delta = 24$ (\autoref{s:eye}). The apparent speed increases to $6.7c$ on parsec scales, as measured by MOJAVE \citep{2019ApJ...874...43L}.
Under these conditions, the gamma-to-radio delay is expected to be $\Delta t = 2$--4~months in observer's frame. This delay is consistent with the observed light curves for \pks and earlier studies by \citet{MOJAVE_XIV,2019ApJ...874...43L,2021ApJ...923...67H}.
Measuring this delay precisely is challenging due to the complex flaring behavior of the blazar.

\section{Summary}

This paper presents a detailed multi-wavelength study of a prominent multi-messenger flare from the quasar \pks, which is temporally associated with the high-energy neutrino IceCube-240105A. Using contemporaneous gamma-ray, X-ray, optical, and radio monitoring, complemented by parsec-scale radio imaging and archival spectral data, we characterize the flare’s timing, energetics, and spectral energy distribution (SED).

We model the broadband SED and interpret the neutrino–electromagnetic flare by introducing a hadronic component and an increase in the Doppler factor within a single-zone framework. The parsec-scale polarization morphology indicates a very small viewing angle, about or below $0.6^\circ$, consistent with the observed apparent jet speed of about $6c$ and SED-fitted  Doppler factors of order 20.

We discover a $90^\circ$ flip in the parsec-scale core linear polarization direction occurring after the neutrino detection. This suggests the emergence or strengthening of a shock emission region. Together with the nearly simultaneous flare starting time from gamma-ray to radio, this supports a scenario in which the high-energy emission originates near the jet base.

Our results indicate that proton-induced interactions within the jet are a plausible mechanism for producing the observed neutrino while remaining consistent with electromagnetic constraints. Overall, we provide strong observational evidence linking the neutrino event to jet activity in \pks (a-posteriory chance coincidence estimated as 2\,\%) and emphasize the importance of continued multi-messenger observations to constrain the underlying emission physics.

\begin{acknowledgements}

We are grateful to the anonymous referee, whose comments have significantly helped us to improve the manuscript.
We thank Egor Podlesnyi for fruitful discussions of the SED modeling, Teresa Toscano and Aditya Tamar for useful comments on the manuscript.

This research was funded by the European Union (ERC MuSES project No~101142396).
A.V.~Plavin is a postdoctoral fellow at
the Black Hole Initiative, which is funded by grants from the John Templeton Foundation (grants 60477, 61479, 62286) and the Gordon and Betty Moore Foundation (grant GBMF-8273). 
The work by A.K.E, P.I.K., Y.A.K., A.V.~Popkov, Y.V.S., and S.V.T.\ is supported in the framework of the State project ``Science’’ by the Ministry of Science and Higher Education of the Russian Federation under the contract 075-15-2024-541.
The views and opinions expressed in this work are those of the authors and do not necessarily reflect the views of these Foundations.

This research made use of the data from the MOJAVE database maintained by the MOJAVE team \citep{2018ApJS..234...12L}, and from MMDC \citep{MMDC}.
The CSS survey is funded by the National Aeronautics and Space Administration under Grant No.~NNG05GF22G issued through the Science Mission Directorate Near-Earth Objects Observations Program.  
ZTF is supported by the National Science Foundation under Grants No.~AST-1440341 and AST-2034437.
This work is partly based on the data obtained with the RATAN-600 radio telescope and Zeiss-1000 and AS-500/2 optical reflectors at the Special Astrophysical Observatory of the Russian Academy of Sciences (SAO RAS).

The National Radio Astronomy Observatory is a facility of the National Science Foundation operated under cooperative agreement by Associated Universities, Inc.
This research has made use of the NASA/IPAC Extragalactic Database, which is funded by the National Aeronautics and Space Administration and operated by the California Institute of Technology.

\end{acknowledgements}

\bibliographystyle{aa}
\bibliography{0446}

\appendix
\section{Electromagnetic observations of \pks, data processing and calibration}
\label{sa:obs}

\begin{figure*}[t]
\centering
\includegraphics[width=0.99\linewidth]{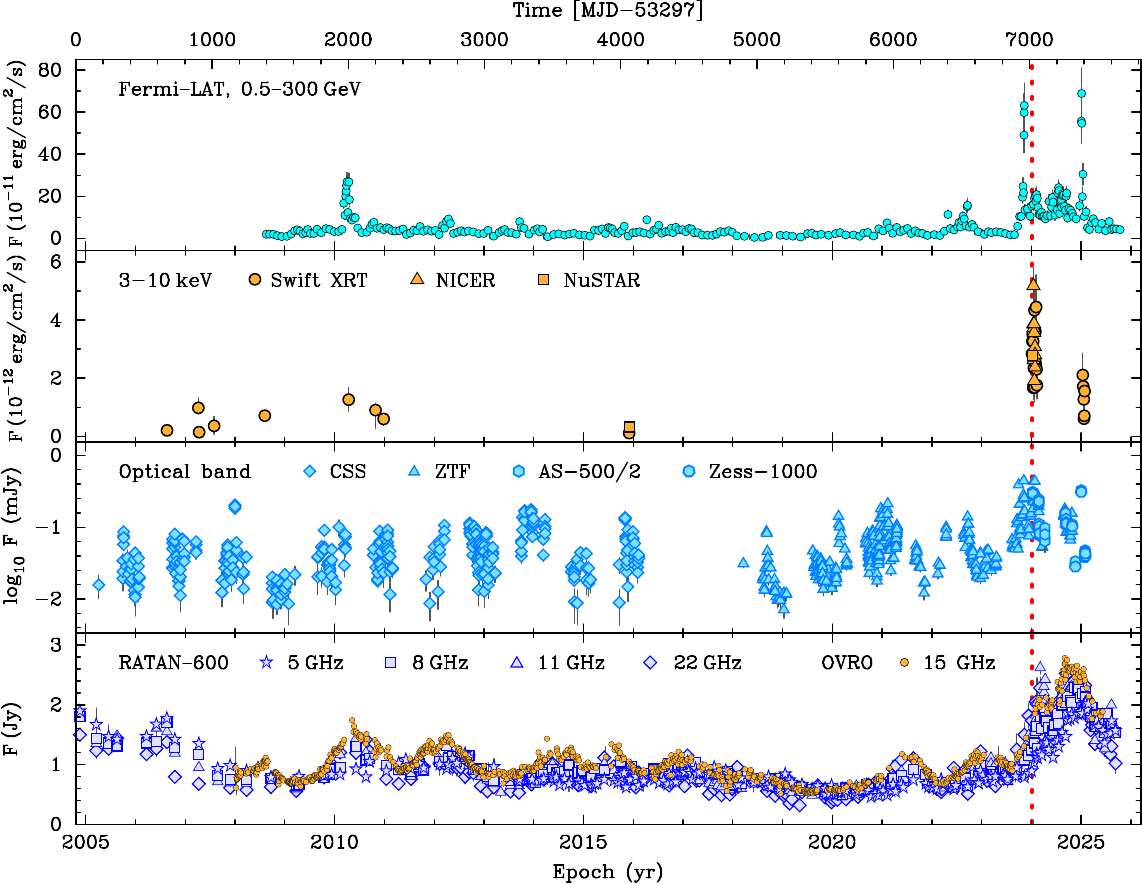}
\caption{The full multi-band light curve for \pks.
See detailed comments in \autoref{f:lc}.
\label{fa:lc}
}
\end{figure*}

The partial (\autoref{f:lc}) and full (\autoref{fa:lc}) multi-messenger light curves of the quasar \pks show the data used in this work. Their observations and calibration are discussed below.

\subsection{Radio band}

\pks is being regularly observed at the RATAN-600 radio telescope \citep{1979S&T....57..324K, 1993IAPM...35....7P} since 1997 as a part of a program of monitoring of VLBI-bright blazars. After the IceCube alert in the beginning of 2024, we started more frequent RATAN-600 observations of this source within the framework of our neutrino trigger program. RATAN-600 is a transit-mode telescope that allows to obtain quasi-simultaneous broad-band radio spectra. During the transit of a source across the meridian in about 5~minutes, flux densities at six frequency bands are measured: 0.96/1.2, 2.3, 4.7, 7.7/8.2, 11, and 22\,GHz; central frequencies of two bands have changed during the monitoring due to receiver upgrades. The data reduction procedure is described in detail in \citet{1999A&AS..139..545K, 2011AstBu..66..109T, 2016AstBu..71..496U, 2018AstBu..73..494T}. 
Flux density calibration was made using the calibration curve covering the declination range between $-35^{\circ}$ and $+49^{\circ}$. It was constructed from observations of a set of calibrators: J0025$-$26, J0137$+$33, J0240$-$23, 
J0627$-$05, J1154$-$35, J1331$+$30 (3C\,286), J1347$+$12, J2039$+$42 (DR\,21), J2107$+$42 (NGC\,7027), and J0542$+$49 (3C 147), see for details \citet{1977A&A....61...99B, 1994A&A...284..331O, 1985AISAO..19...60A, 2013ApJS..204...19P, 2017ApJS..230....7P}. The characteristic cadence of the RATAN-600 observations of \pks ranged from one to several months before the IceCube alert and from one day to two weeks after it. 
In the paper, the RATAN-600 frequencies are denoted by their rounded values: 5, 8, 11, and 22~GHz. We do not use the data at 1 and 2~GHz in our analysis because they are affected by radio interference. The trigger program data from 2024-2025 were averaged in approximately two-weak bins for improving the signal-to-noise ratio.

The Owens Valley Radio Observatory (OVRO) 40-meter Telescope has been dedicated to monitoring a sample of $\sim 1830$ blazars with a typical cadence of 3-4~days since January 2008. It uses dual-beam optics straddling the central focus point in double switching mode to minimize atmospheric effects.
The cryogenic receiver is centered at 15~GHz with a 2~GHz equivalent noise bandwidth.
A temperature stabilized noise diode is used to compensate for gain drifts.
The quasar 3C\,286 is used for absolute flux density calibration, with an assumed value of 3.44~Jy \citep{1977A&A....61...99B}. Occasionally, the molecular cloud DR\,21 is used, calibrated relative to 3C\,286 using early OVRO 40~m data. 
The details of the observing method, observations and data reduction are presented by \citet{2011ApJS..194...29R}. 

The MOJAVE program (Monitoring Of Jets in Active galactic nuclei with VLBA Experiments) performs the longest monitoring of active galactic nuclei (AGN) by the VLBA at 15~GHz, producing the highest number of source epochs among all similar projects. This program started in 1994 with Stokes I observations, and since 2002 includes polarization measurements.
See \citet{2018ApJS..234...12L} and references therein.
Completely calibrated data in the form of visibility and image FITS files are made available in an online archive\footnote{\url{https://www.cv.nrao.edu/MOJAVE/}}.
We note that typical MOJAVE EVPA (Electric Vector Position Angle) uncertainties are estimated as 2-3~deg. In this paper we present and discuss EVPA values for the core region (\autoref{f:lc}), deduced from MOJAVE-calibrated VLBA images (\autoref{f:MOJAVE_pol}). We adopt more conservative error estimate of 5~deg for the core EVPA.
The quasar \pks was originally observed for two epochs in 1998 as part of the project BG077, which targeted high redshift blazars.
In 2002--2010 it was observed as a member of the complete MOJAVE sample with a main goal to determine jet kinematics \citep{2019ApJ...874...43L}. 
Its observations restarted in 2022, in support of MOJAVE-neutrino studies. 

\subsection{Optical band}

Optical photometric data were obtained from two publicly available surveys. 
The Catalina Sky Survey (CSS; \citealt{Drake2009})\footnote{\url{https://catalina.lpl.arizona.edu/}} provides $V$-band light curves 
covering the period from 5~April 2005 to~9 February 2016. 
The original CSS magnitudes ($V_\mathrm{css}$) were transformed into the Johnson $V$ system 
using the relation given by \citet{Drake2009}
$
    V = V_\mathrm{css} + 0.31\,(B-V)^2 + 0.04\,,
$
where the color index $B-V$ was estimated from the median ZTF color 
$\langle g-r \rangle = 0.983$ by applying the relation from \citet{Jester2005}:
$
    g-r = 0.93\,(B-V) - 0.06\,.
$
The Zwicky Transient Facility (ZTF; \citealt{Masci2019})\footnote{\url{https://www.ztf.caltech.edu/index.html}} contributes 
$g$- and $r$-band photometry between 20 March 2018 and 26 October 2024 (DR23).

The optical study (mainly in the $R$-band) was carried out with the 0.5-metre AS-500/2 (January 2024 -- September 2024) and 1-metre Zeiss-1000 (October 2024 -- January 2025) optical reflectors of SAO RAS.

The observations with the optical telescope Zeiss-1000 were conducted with a CCD photometer in the Cassegrain focus, which is equipped with a $2048\times2048$~px back-illuminated E2V~chip CCD\,42-40. Details of the instrumental setup are described in \citet{2023AstBu..78..464V}. 
The main characteristics of the instrumental complex of an AS-500/2 reflector are presented in \citet{2022Photo...9..950V}.
In order to improve the system parameters, we have installed a smaller but more sensitive detector in July 2023: the back-illuminated electron-multiplying CCD camera Andor IXon$^{\rm EM}$+897 with 512$\times$512 px chip.
The typical integration time for observations of the blazar was 300~s for Zeiss-1000 and 90~s for AS-500/2. The optical data are collected from 14 observing nights between January 2024 and January 2025.

\subsection{X-ray band}

X-ray monitoring data were collected from multiple public archives. 
For the quiescent state, we used Swift-XRT and NuSTAR data retrieved from the public data collections (see \autoref{s:obs_public}). 
Within the flare period (MJD 60315--60351; 2024 January 6 to February 11), additional observations from Swift-XRT, NICER, and NuSTAR were included. 
The fluxes were homogenized to the 3--10 keV energy band and are shown in \autoref{f:lc}.  

NuSTAR data were taken from the Astronomer’s Telegram report of \citet{Woo2024} and converted to the 3--10 keV band. 
NICER data products were retrieved from the NICERMASTR database \citep{nicer}, and Swift-XRT products from the SWIFTMASTR database \citep{SwiftMC}. 
Swift-XRT and NICER spectra were fitted in \texttt{XSPEC} \citep{xspec} with an absorbed power-law model 
(\texttt{tbabs*powerlaw}), adopting a fixed Galactic hydrogen column density of 
$N_\mathrm{H} = 2.48 \times 10^{20}\,\mathrm{cm}^{-2}$. 
No significant variability was detected across the flare epochs. 
The weighted mean flux in the 3--10~keV band is 
$(3.11 \pm 0.25) \times 10^{-12}\,\mathrm{erg\,cm^{-2}\,s^{-1}}$ 
with a photon index $\Gamma = 1.06 \pm 0.05$. These values are consistent with an independent analysis reported in Astronomer's Telegrams: 
\citet{Sharpe2024} obtained from NICER an average photon index of 
$\Gamma = 1.34 \pm 0.25$ and an average unabsorbed flux of 
$(4.76 \pm 1.32) \times 10^{-12}\,\mathrm{erg\,cm^{-2}\,s^{-1}}$ in the 0.2--12 keV range, 
while \citet{Prince2024} reported Swift-XRT fluxes of 
$(4.0 \pm 1.0) \times 10^{-12}$ and $(4.4 \pm 0.7) \times 10^{-12}\,\mathrm{erg\,cm^{-2}\,s^{-1}}$ 
on January 6 and 8, respectively, with photon indices $\Gamma = 1.4 \pm 0.5$ and $\Gamma = 1.21 \pm 0.22$.  

For SED modeling during the flare, we used one representative NICER epoch with the longest exposure 
(MJD 60327; 2024 January 18; ObsID 6736120108; exposure time 8996~s) rather than a combination of all epochs to avoid possible systematics due to imperfect background subtraction. 
We derived unabsorbed fluxes in four energy bins (0.3--1, 1--2, 2--5, and 5--10~keV; \autoref{f:SED}, left). 
NICER results from \citet{Sharpe2024} are also included in the same plot and are in good agreement with our measurements.  

\subsection{Gamma-ray band}
\label{s:gamma}

To construct the gamma-ray light curve, we applied the adaptive binning \citep{Lott12} to the data of the \textit{Fermi} Large Area Telescope (LAT). This technique enables to reveal more details compared to fixed-binning, especially during bright flares, and allows to create a light curve with a constant relative flux uncertainty, which we set at the level of 20\%. We used the current version of the Fermitools software package, v.2.2.0. A source model was produced by the script {\sc make4FGLxml.py} setting the 4FGL-DR3 source catalog, \textsc{gll\_iem\_v07.fits} as a Galactic interstellar emission model designed for point source analysis, Pass 8 Source front+back class events, and \textsc{iso\_P8R3\_SOURCE\_V3\_v1.txt} for the isotropic spectral template. The Region-of-Interest (RoI) around the target source was set to $20^\circ$ and accumulated 123 point-like sources and no extended ones. The procedure {\sc gtselect} produced data cuts within 0.1--300~GeV energy range and from 2008-08-04 to 2025-10-02 time range. Adaptive binning was performed using a normal-time arrow and assuming that the source spectrum is constant with time. To obtain more accurate fluxes of the target source, the fluxes of two bright sources in the RoI were taken into account. One of these sources is positionally associated with the quasar 0506+056 separated by $7\fdg6$ from 0446+112. The bin widths derived are distributed from about 1 day during the flares in 2024 and 2025 to 70 days for the low-state periods, with a median of 21 days.

\section{A Bayesian block analysis of flares in the Fermi LAT light curve of \pks}\label{sa:flare}

\begin{figure}
\centering
\includegraphics[width=1.00\linewidth]{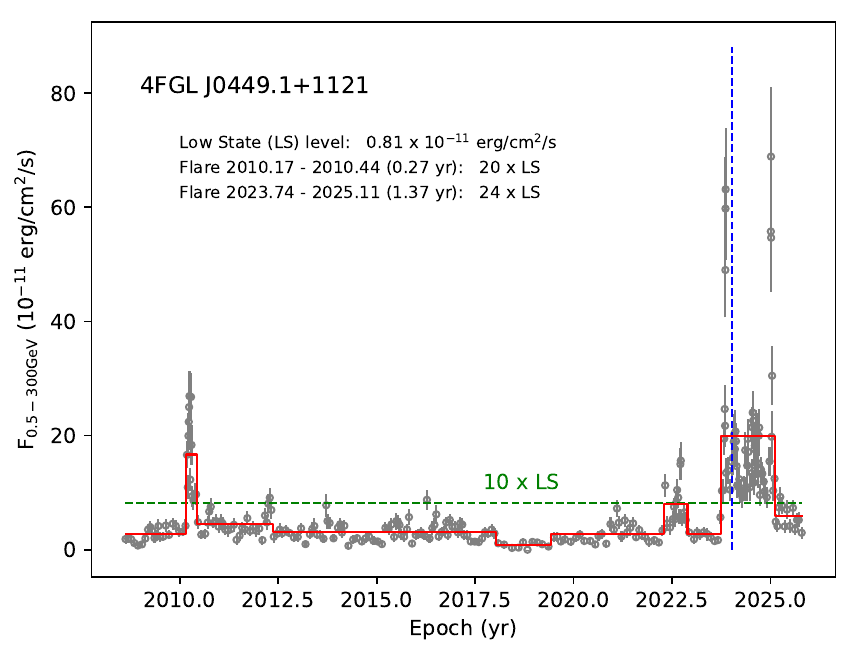}
\caption{
The \textit{Fermi} LAT light curve of \pks segmented by Bayesian blocks (red broken line). Two flares are identified, each exceeding 10 times the level of the low state (green dashed line). The neutrino event epoch is marked by the blue dashed line.
}
\label{fa:bb}
\end{figure}

Following a non-parametric modeling technique developed by \cite{2013ApJ...764..167S}, we divided the {\it Fermi} LAT data into ten Bayesian blocks (\autoref{fa:bb}) with a value of false positive probability $p_0 = 5.7\times10^{-7}$, equivalent to about  $5\sigma$ detection threshold in Gaussian statistics. This conservative value is typically adopted for structure analysis of light curves in high-energy astrophysics to robustly isolate flares or quiescence \citep[e.g.,][]{2016ApJ...829....7L}. We define the quiescent background level as a minimum flux among the derived Bayesian blocks, assuming that flare tails do not make a significant contribution to it. In this way, the low state (LS) is $0.85\times10^{-11}$~erg/cm$^2$/s, reflecting the lowest source activity level. Using traditional determination of a flare as 10 times flux increase of a low state level, we select two flares with all data points above the flare level: 2010.166--2010.440 (100~d duration, $20.5\times\text{LS}$) and 2023.744--2025.113 (500~d duration, $24.5\times\text{LS}$). There are two short spikes embracing the recent long flare. The durations of the left and right spikes are about 6~d and 11~d, respectively. Note that the neutrino event is not coincident with the left spike, it is offset by 58~d.

\end{document}